\documentclass[12pt]{article}

\newcommand{\x}{arXiv:}
\newcommand{\m}{\mathrm}
\newcommand{\be}{\begin{equation}}
\newcommand{\ee}{\end{equation}}
\newcommand{\ba}{\begin{eqnarray}}
\newcommand{\ea}{\end{eqnarray}}

\usepackage{graphicx}
\usepackage{amssymb}
\usepackage{amsmath}

\begin{document}
\thispagestyle{empty}
\begin{center}

\null \vskip-1truecm \vskip2truecm

{\Large{\bf \textsf{The Trajectory of the Cosmic Plasma }}}

{\Large{\bf \textsf{Through the Quark Matter Phase Diagram}}}

{\large{\bf \textsf{}}}

\vskip1truecm

{\large \textsf{Brett McInnes}}

\vskip1truecm

\textsf{\\  National
  University of Singapore}

\textsf{email: matmcinn@nus.edu.sg}\\

\end{center}
\vskip1truecm \centerline{\textsf{ABSTRACT}} \baselineskip=15pt
\medskip

Experimental studies of the Quark-Gluon Plasma (QGP) focus on two, in practice distinct, regimes: one in which the baryonic chemical potential $\mu_B$ is essentially zero, the other in which it is of the same order of magnitude as the temperature. The cosmic QGP which dominates the early Universe after reheating is normally assumed to be of the first kind, but recently it has been suggested that it might well be of the second: this is the case in the theory of ``Little Inflation''. If that is so, then it becomes a pressing issue to fix the trajectory of the Universe, as it cools, through the quark matter phase diagram: in particular, one wishes to know where in that diagram the plasma epoch ends, so that the initial conditions of the hadronic epoch can be determined. Here we combine various tools from strongly coupled QGP theory (the latest lattice results, together with gauge-gravity duality) in order to constrain that trajectory, assuming that Little Inflation did occur.

\newpage
\addtocounter{section}{1}
\section* {\large{\textsf{1. The Cosmic Plasma in the Quark Phase Diagram}}}
The history of the early Universe is conventionally divided into various ``epochs'': the Planck epoch, the plasma epoch, and so on. Each epoch is described by a number of parameters, and it is the task of early Universe physics to determine the trajectory of the cosmos, as time passes, through these parameter spaces.

The plasma epoch, just prior to the more familiar hadronic epoch, is the earliest one involving forms of matter that can be produced experimentally, and it is therefore of particular interest to understand it more completely. During that period, the Universe consisted predominantly of a Quark-Gluon Plasma (QGP), and the relevant parameter space is the quark matter phase diagram \cite{kn:itoh,kn:ohnishi,kn:mohanty,kn:satz}, in which the parameters are temperature $T$ and the baryonic chemical potential $\mu_B$. This diagram is thought to take the form shown in Figure 1; as usual in such diagrams, there is a line along which there is a first-order phase transformation (to the hadronic or other, more exotic states at very large $\mu_B$) but this line does not persist to low values of $\mu_B$: instead it terminates at a \emph{critical endpoint} \cite{kn:race}, to the left of which there is a smooth crossover; this has been known for some time, as a result of highly sophisticated lattice QCD computations \cite{kn:aoki}.

As the Universe cooled, the plasma traced out a trajectory downwards through this diagram. This immediately prompts a basic question: as the cosmic plasma cooled, \emph{did it undergo an actual phase transition, or did it merely pass through a smooth crossover?} The answer to this question is of basic importance in understanding the initial conditions for the hadronic epoch \cite{kn:raf1}.

\begin{figure}[!h]
\centering
\includegraphics[width=0.85\textwidth]{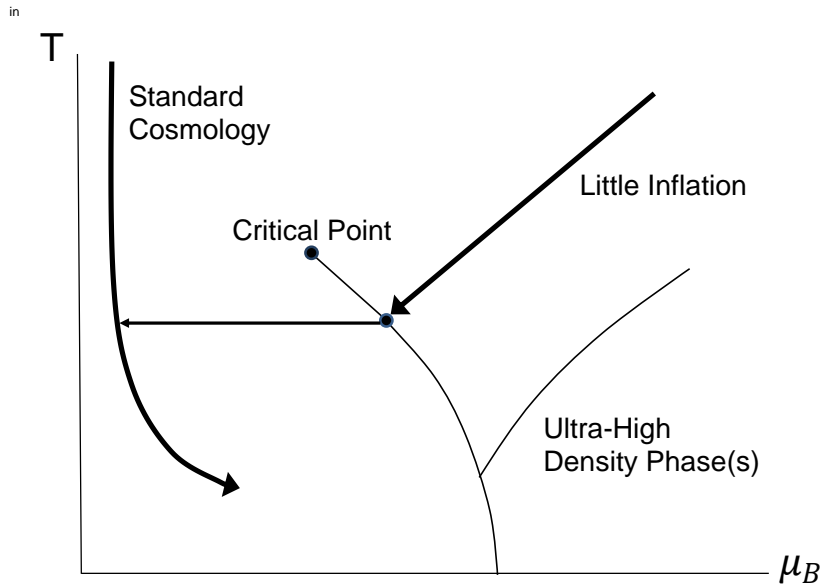}
\caption{Possible Trajectories of the cosmic plasma in the Quark Matter Phase Diagram (after Boeckel and Schaffner-Bielich \cite{kn:tillmann3}) }
\end{figure}

The standard (current) answer to this question is that, in view of the small value ($\eta_B \approx 10^{-9}$) of the net baryon density/entropy density ratio in the hadronic epoch, $\mu_B$ must have been small (relative to the temperature) through the plasma epoch; so the evolution proceeded in a continuous way, through a smooth crossover. (On the quark matter phase diagram, the evolution is represented by a trajectory which is almost vertical, and very close to the $T$ axis, until the temperature falls far below the critical temperature; at around 35 MeV it then deviates sharply to the right, eventually entering the region corresponding to nuclear matter. See \cite{kn:conv}.) This has not always been the consensus view, however: in \cite{kn:wit}, Witten considered the possible observational signatures of a cosmic first-order phase transition, and in \cite{kn:ng} and \cite{kn:borg} the possibility that the cosmic plasma era ended at a large value of $\mu_B$ was considered\footnote{Large values of $\mu_B$ also play a role in the theory of ``inhomogeneous baryogenesis'': see \cite{kn:silk,kn:dolgov}.}.

More recently, in \cite{kn:tillmann1,kn:tillmann2,kn:tillmann3}, this idea has been revived and developed much more extensively: the result is a theory of ``\emph{Little Inflation}''. One of the principal motivations of this theory concerns the well-known Affleck-Dine theory of baryogenesis \cite{kn:dine} (see also \cite{kn:linde}). This theory can indeed produce the observed value of $\eta_B$, but \cite{kn:dinereview} it can also easily produce \emph{substantially larger} values. The point of view of Little Inflation is that one might regard (relatively) large values of $\mu_B$, arising from Affleck-Dine baryogenesis or some related mechanism, to be \emph{generic} for the cosmic plasma. In fact, Little Inflation can handle, and is consistent with, quite large values of $\mu_B$ (relative to the temperature $T$), perhaps as large as $\mu_B/T \approx 100$.

In this theory, then, the value of $\mu_B$ was large throughout the plasma epoch, even when it ended, through a first-order transition. The theory postulates that, as the plasma crossed the phase transition line, it first entered a false (metastable) QCD vacuum, where it was trapped and inflated under the influence of the constant potential. (The number of e-folds is less than 10, hence ``Little'' Inflation.) When the plasma transits to the true vacuum, the corresponding reheating dilutes the baryon/photon ratio and the value of $\eta_B$ drops rapidly to around 10$^{-9}$. (On the quark matter phase diagram, this means that the trajectory of the plasma is a diagonal line which descends towards, and intersects, the phase line. After that, the trajectory is nearly horizontal, until it meets, and merges with, the conventional evolution at some point just below the critical temperature ---$\,$ in fact, at essentially the same temperature as the point where it met the phase line. See Figure 1 (based on Figure 2 of \cite{kn:tillmann3}).

This scenario is in good agreement with the data, and makes numerous predictions for forthcoming observations. In particular, it makes definite predictions regarding the extremely intense magnetic fields which are now thought to have permeated the cosmic plasma, \emph{and which must be taken into account in any theory of that plasma}: see \cite{kn:reviewA}\cite{kn:reviewB} for extensive recent reviews. That is, Little Inflation automatically incorporates a ``magnetogenesis'' mechanism. The magnetic field is an essential component of Little Inflation and must be included in any (for example, holographic) analysis of it.

From a theoretical point of view, this way of completing theories of baryogenesis is very attractive, since the small observed value of $\eta_B$ is accounted for by the dramatic changes which, \emph{in any case}, must occur if the plasma era reaches its end through a first-order phase transition.

Thus, we have two very different accounts of the manner in which the cosmic plasma traverses the quark matter phase diagram. It is remarkable that in these theories the evolution of the cosmic plasma precisely reflects the QGP as it is produced in two very different current programmes of heavy-ion collision experiments: on the one hand, the low-$\mu_B$ plasma is being examined by ALICE experiment at the LHC \cite{kn:andronicover}, while the high-$\mu_B$ plasma has been the target of the beam scan experiments at the RHIC \cite{kn:ilya,kn:dong,kn:STAR}, and will be further explored at SHINE, NICA and FAIR, and the extended beam scan at the RHIC \cite{kn:shine,kn:nica,kn:fair,kn:BEAM}. Although comparing the cosmic plasma with the QGP produced in heavy-ion collisions is far from straightforward \cite{kn:raf2}, it is reasonable to hope that these experiments might yield at least some data relevant to the cosmic plasma near the end of the plasma epoch: one set of experiments will be relevant to the conventional picture of the evolution of the cosmic plasma, the other set to Little Inflation.

If the hadronic epoch began after an actual first-order phase transition, this could have important consequences. To take but three of many examples:

$\bullet$ A sharp phase transition could give rise to inhomogeneities that might help to explain currently observed structures, such as voids \cite{kn:raf2}.

$\bullet$ There have been suggestions \cite{kn:shur} that a first-order cosmic phase transition might be accompanied by acoustic effects, giving rise to observable signatures in the form of gravitational waves.

$\bullet$ A large value of $\mu_B$ during the plasma epoch could modify the plasma equation of state in potentially observable ways \cite{kn:sanches}.

Little Inflation does however suffer from one drawback: unlike the conventional theory, it does not specify precisely the actual trajectory of the cosmic plasma through the phase diagram. Thus, in particular, one does not have a prediction for the location of the point on the phase line where the plasma epoch ended; that is, the theory does not fix initial conditions for the hadronic epoch. \emph{Our objective here is to ameliorate this problem}.

We do this in two steps. First, the latest results from lattice gauge theory \cite{kn:karsch} are used to constrain the position of the phase transition line; and then gauge-gravity duality (see \cite{kn:veron} for the general case, \cite{kn:youngman,kn:gubser,kn:janik} for the application to the QGP) is used to impose a lower bound on the slope of the Little Inflation trajectory shown in Figure 1, in the manner explained in \cite{kn:82}\cite{kn:83}. We argue that, as the nucleation of a false vacuum is a delicate process requiring that specific conditions be satisfied regarding the height and the stability of the relevant barrier ---$\,$ see section III of \cite{kn:tillmann3} ---$\,$ it is essential to avoid, as far as possible, the vicinity of the critical endpoint. For one expects the typical strong fluctuation phenomena associated with \emph{critical opalescence} \cite{kn:csorgo} to be important in that region of the phase diagram; indeed, those fluctuations are so strong that they provide the principal way of locating the critical endpoint \cite{kn:STARCRIT,kn:PHENIX,kn:turko}. Combining all of these arguments, we can constrain the trajectory of the cosmic plasma rather strongly.

\addtocounter{section}{1}
\section* {\large{\textsf{2. The Phase Transition Line}}}
There is as yet no complete agreement as to the location of the quark matter critical endpoint, as obtained from lattice studies: in \cite{kn:karsch} one finds a discussion of various proposed points in the phase diagram, such as one (see \cite{kn:gavai}) with coordinates $(\mu_B^{CEP}/T^{CEP}, T^{CEP}/T^C_0) \approx (1.68, 0.94)$, where $T^C_0$ is the crossover temperature at $\mu_B = 0$, for which we follow \cite{kn:karsch} and take the value $\approx$ 150 MeV; but also another with coordinates
$(\mu_B^{CEP}/T^{CEP}, T^{CEP}/T^C_0) \approx (2.2, 0.99)$ (see \cite{kn:fodor}). In \cite{kn:race}, a range for $\mu_B^{CEP}/T^{CEP}$ of $1 - 2$ is suggested. However, it is argued in \cite{kn:karsch} that recent lattice studies of the equation of state at non-zero $\mu_B$, and of cumulants of charge fluctuations, both indicate that a critical endpoint satisfying $\mu_B^{CEP}/T^{CEP} < 2$ is ``unlikely''. In view of all this, let us put the critical endpoint at $T^{CEP}$ = 145 MeV, $\mu_B^{CEP}$ = 300 MeV, since that is a point on the curve shown in Figure 6 of \cite{kn:karsch}. We stress, however, that the question of the location of the critical point is far from settled: we are selecting $T^{CEP}$ = 145 MeV, $\mu_B^{CEP}$ = 300 MeV simply to be definite, \emph{not} to assert that these values are confirmed.

The shape of the phase line\footnote{Here and throughout we are assuming, as is customary (but see the discussion in \cite{kn:souza}), that the phase line corresponds to the breaking of chiral symmetry \cite{kn:pisar}.} has, again, recently been constrained by lattice investigations: according to \cite{kn:kacz} (very recently confirmed in \cite{kn:karsch}), the (absolute value of the) curvature of the phase line is, at least for small $\mu_B$, given approximately by
\begin{equation}\label{A}
\kappa_B = 0.0066(7),
\end{equation}
where $\kappa_B$ is expressed in terms of the baryonic rather than the quark chemical potential\footnote{There has in fact been some very recent debate regarding this value: see \cite{kn:cea,kn:bonati1,kn:bonati2,kn:new,kn:newer}. Again, our concern here is to be definite, not to take a position on this question; in any case, the differences are not such as would change our main conclusions substantially.}. Notice that this curvature is very small, far smaller than is represented in the usual schematic quark matter phase diagrams (including Figure 1 above). The corresponding curve, reasonably extrapolated, is shown in Figure 6 of \cite{kn:karsch}.

In the application to Little Inflation, the trajectory of the cosmic plasma intersects this line at some point to the right of the critical endpoint. We now try to use gauge-gravity duality to constrain the location of this intersection point.

\addtocounter{section}{1}
\section* {\large{\textsf{3. The Constraint from Gauge-Gravity Duality}}}
As the phase transition line bends down and to the right from the critical endpoint (see Figure 1) it follows that, if indeed a given plasma passes across this line, then the value of $\mu_B/T$ at the point where the trajectory crosses the phase line, $\mu_B^{\#}/T^{\#}$, must satisfy
\begin{equation}\label{B}
\mu_B^{CEP}/T^{CEP}\;\leq\;\mu_B^{\#}/T^{\#};
\end{equation}
so we have a lower bound on $\mu_B^{\#}/T^{\#}$. We now recall \cite{kn:83} that gauge-gravity duality imposes an \emph{upper} bound on it.

To see how this works, consider the submanifold of anti-de Sitter spacetime AdS$_4$ which is foliated by conformal copies of three-dimensional Minkowski spacetime. Its metric may be expressed as
\begin{eqnarray}\label{C}
g(\m{AdS_4^0}) \;=\; -\, {r^2\over L^2}\text{d}t^2\;  + \;{\text{d}r^2\over r^2/L^2} \;+\;r^2\left[\text{d}\psi^2\;+\;\text{d}\zeta^2\right],
\end{eqnarray}
where $L$ denotes the curvature scale, and the zero superscript reminds us that the spacetime is foliated by (1+2)-dimensional zero-curvature sections. We can regard $t$ as time at infinity, and $\psi$ and $\zeta$ as dimensionless planar coordinates there. Clearly the geometry at infinity is indeed that of (1+2)-dimensional Minkowski spacetime, if we make a conformal transformation using the factor $r^2/L^2$.

We ask the reader to regard this (1+2)-dimensional spacetime as a timelike section through ordinary four-dimensional Minkowski spacetime: that is, one constructs it by arbitrarily choosing a specific spatial direction in three-dimensional flat space, described by a dimensionless coordinate $\xi$, and regarding $t$, $\psi$, and $\zeta$ as coordinates in the (1+2)-dimensional spacetime perpendicular to that direction (so that the full four-dimensional Minkowski spacetime has metric $-\text{d}t^2\;+\;L^2\left[\text{d}\psi^2\;+\;\text{d}\zeta^2\;+\;\text{d}\xi^2\right]$). That is, the conformal infinity of (this version of) AdS can be regarded as a (1+2)-dimensional sub-spacetime of four-dimensional Minkowski spacetime.

Now the conformal factor we used here is not the only possible choice: we are free to make a further transformation involving $t$ as well as $r$. Let $a(t)$ be any smooth function of $t$, let $\tau$ be defined by $\text{d}\tau = a(t)\text{d}t$, and use this to express $t$ as a function of $\tau$; then let $A(\tau)$ denote $a(t)$ when this is done. A conformal transformation of the flat spacetime, with conformal factor $a(t)^2$, produces the conformally flat FRW metric $-\text{d}\tau^2\;+\;A(\tau)^2L^2\left[\text{d}\psi^2\;+\;\text{d}\zeta^2\;+\;\text{d}\xi^2\right]$. So we see that the conformal infinity of AdS$_4$ can be regarded as a (1+2)-dimensional sub-spacetime of any spatially flat FRW spacetime.

Now of course FRW spacetimes have a very large isometry group: they are homogeneous and isotropic. In a physical sense, a four-dimensional spatially flat FRW spacetime is ``really'' (1+2)-dimensional: for if we understand the physics on any two-dimensional plane, then we automatically know the conditions on any other two-dimensional plane, even if it has a different orientation and location. In short, it suffices to understand the physics associated with a (1+2)-dimensional sub-spacetime with metric $-\text{d}\tau^2\;+\;A(\tau)^2L^2\left[\text{d}\psi^2\;+\;\text{d}\zeta^2\right]$.

As we have seen, this spacetime is holographically dual to AdS$_4$, and so we have a way of studying this FRW spacetime holographically, section by section. This duality is of little interest, however, since for example AdS$_4$ does not have a well-defined temperature, or any parameter corresponding to the baryonic chemical potential of the boundary theory; nor does it allow for the strong magnetic fields which, as we discussed earlier, are thought to have permeated the cosmic plasma. However, there is a very well-known way of solving these problems \cite{kn:youngman,kn:gubser,kn:janik}: instead of using AdS$_4$, let us insert an asymptotically AdS dyonic Reissner-Nordstr\"om metric with planar (zero-curvature) event horizon into the bulk. The metric, an exact solution of the Einstein-Maxwell equations \cite{kn:lemmo}\cite{kn:dyon}, is
\begin{eqnarray}\label{D}
g(\m{AdSdyRN^{0}_{4})} & = & -\, \Bigg[{r^2\over L^2}\;-\;{8\pi M^*\over r}+{4\pi (Q^{*2}+P^{*2})\over r^2}\Bigg]\m{d}t^2\; \nonumber \\
& &  + \;{\m{d}r^2\over {\dfrac{r^2}{L^2}}\;-\;{\dfrac{8\pi M^*}{r}}+{\dfrac{4\pi (Q^{*2}+P^{*2})}{r^2}}} \;+\;r^2\Big[\m{d}\psi^2\;+\;\m{d}\zeta^2\Big],
\end{eqnarray}
where $M^*$, $Q^*$, and $P^*$ are geometric parameters connected to the mass and electric and magnetic charges (per unit horizon area), $\psi$ and $\zeta$ are dimensionless Cartesian coordinates, and $L$ is now the \emph{asymptotic} AdS curvature parameter. This is the minimally complex bulk metric that allows us to account for a dual field theory which [1] has a well-defined temperature, [2] has a non-zero baryonic chemical potential, and [3] is accompanied by a non-zero magnetic field, as discussed earlier.

The electromagnetic potential form outside the black hole is
\begin{equation}\label{BLUE}
A\;=\;\left ({1\over r_h}\,-\,{1\over r}\right ){Q^*\over L}dt\;+\;{P^*\over L}\psi d\zeta,
\end{equation}
where the constant term in the coefficient of $dt$ is present to ensure regularity; here $r_h$ represents the location of the event horizon. The field strength form is
\begin{equation}\label{CHEAP}
F\;=\;-\,{Q^*\over r^2L}dt \wedge dr \;+\;{P^*\over L}d\psi \wedge d\zeta.
\end{equation}
The baryonic chemical potential $\mu_B$ of the dual system is related holographically to the asymptotic value of the time component of the potential form, that is, to $Q^*/(r_hL)$, while the magnetic field associated with the dual system is given by the asymptotic value of the field strength, that is, to $P^*/L$; see below for the precise relations. This procedure, making use of dyonic AdS black holes as duals of strongly coupled systems with large chemical potentials and subjected to strong magnetic fields, is standard in applications of holography to condensed matter physics (see \cite{kn:hartkov}), and it works in just the same way here.

Now the conformal boundary here is \emph{precisely the same} as in the much simpler case of AdS$_4$, discussed above: it is conformal to a (1+2)-dimensional section of four-dimensional Minkowski spacetime. This bulk geometry is therefore dual, in precisely the same sense as we discussed earlier, to physics on some (1+2)-dimensional section through a FRW spacetime with flat spatial sections; but now the bulk does have a well-defined temperature (the Hawking temperature of the black hole), as well as having additional parameters ($Q^*$ and $P^*$) which can describe the baryonic chemical potential $\mu_B$ and the magnetic field $B$ in a plasma at infinity. Thus we obtain a gauge-gravity description of certain quantities characterizing the cosmic plasma. (For a detailed description of this procedure, see \cite{kn:82,kn:83}.)

Of course, this duality is severely limited, since the bulk is static, and so we can only hope to account for (approximately) time-independent quantities on the boundary. Such quantities include two of particular interest, however: namely the temperature-scaled magnetic field $\beta$, defined by
\begin{equation}\label{E}
B\;=\;\beta \, T^2,
\end{equation}
and the ``specific baryonic chemical potential'' $\varsigma_B$, defined by
\begin{equation}\label{F}
\mu_B = \varsigma_B \, T.
\end{equation}
These two quantities are constant because $B$ and $T^2$ evolve, in conventional cosmology, at the same rate (with the inverse square of the cosmological scale factor) and similarly $\mu_B$ and $T$ both evolve with the reciprocal of the scale factor; see \cite{kn:83} for the details. Note that $\varsigma_B$ is the reciprocal of the slope of the straight line trajectory\footnote{The straightness of this trajectory should really be understood in a piecewise sense, since the slope changes abruptly at particular instants in the history of the cosmic plasma: see \cite{kn:raf2}. However, we are mainly concerned here with the final segment, terminating at the intersection with the phase line.} shown in Figure 1. These two constants will be our main concern in this work.

The holographic ``dictionary'' relates the black hole parameters to the temperature, baryonic chemical potential, and magnetic field strength of the boundary field theory by means of the equations \cite{kn:83}
\begin{equation}\label{G}
\mu_B\;=\;{3Q^*\over r_hL},
\end{equation}
\begin{equation}\label{H}
B \;=\; P^*/L^3,
\end{equation}
\begin{equation}\label{I}
T\;=\;{1\over 4\pi r_h}\,\Bigg({3r_h^2\over L^2}\;-\; {4\pi (Q^{*2}+P^{*2}) \over r_h^2}\Bigg);
\end{equation}
here $r_h$ is, as above, the value of the radial coordinate at the event horizon, but one can think of it as fixed function of the boundary parameters by combining these equations to obtain
\begin{equation}\label{J}
3r_h^4\,-\,4\pi TL^2r_h^3\,-\,{4\pi \over 9}\mu_B^2L^4r_h^2\,-\,4\pi B^2L^8\;=\;0.
\end{equation}
That is, on the boundary we \emph{define} $r_h$ to be the largest real solution of this equation.

Now it has recently been emphasised, particularly by Ferrari and Rovai \cite{kn:ferrari3}, that the very existence of a holographic duality between a gravitational theory in the bulk and a non-gravitational boundary theory can impose highly non-trivial conditions on both. In particular, it was found in \cite{kn:ferrari3} that holography imposes an ``isoperimetric inequality'' on the (Euclidean) bulk geometry, which runs as follows.

Consider a hypersurface $\Sigma$ in the bulk which is homologous to the three-dimensional boundary, and denote the area of a fixed compact domain in $\Sigma$ by $A(\Sigma)$. Let  $V(M_{\Sigma})$ be the volume enclosed by $\Sigma$. Then the requirement is that (with ``E'' for ``Euclidean'')
\begin{equation}\label{K}
\mathfrak{S^{\m{E}}}\;\equiv\;A(\Sigma)\;- \;{3 \over L}V(M_{\Sigma}) \;\geqslant \;0,
\end{equation}
where, as above, $L$ denotes the asymptotic curvature scale. This inequality is certainly not satisfied in all cases, so we obtain a useful constraint just from the internal consistency of holography.

There is of course a Lorentzian interpretation of the isoperimetric inequality: in the case of a black hole bulk, the Lorentzian version\footnote{The Lorentzian version of (\ref{K}) usually differs from the Euclidean version, because the geometry of the Lorentzian bulk differs from its Euclidean counterpart: one has to complexify not just $t$ but also some other parameters, such as the electric charge and the angular momentum.} of (\ref{K}) states that the black hole should be free of an instability which can result when branes are created in the bulk by a Schwinger-type pair-production effect first studied by Seiberg and Witten \cite{kn:seiberg}: see \cite{kn:maldacena, kn:KPR, kn:SPONT} for general studies of this important effect, and \cite{kn:84} for the specific application to this case. This too is a consistency condition, in the sense that, without it, it is not consistent to assume that the bulk is static or that string-theoretic objects (such as branes) in the bulk can be neglected. Henceforth we focus on the Lorentzian version of consistency.

Using equation (\ref{C}), we can readily evaluate $\mathfrak{S^{\m{E}}}$ and its Lorentzian counterpart $\mathfrak{S^{\m{L}}}$ for AdS$_4^0$: one finds that $\mathfrak{S^{\m{L}}}(\m{AdS^{0}_{4}}) = 0$, and so the consistency condition is satisfied. But for the black hole with metric (\ref{D}), one has \cite{kn:83}
\begin{equation}\label{L}
\mathfrak{S^{\m{L}}}(\m{AdSdyRN^{0}_{4}})(r)\;=\; { \left (-8\pi M^* + {4\pi (P^{*2}+Q^{*2})\over r} \right )/L\over 1+\sqrt{1-{8\pi M^*L^2\over r^3}+{4\pi (P^{*2}+Q^{*2})L^2 \over r^4}}}+{r_h^3\over L^3},
\end{equation}
where, from the definition of $r_h$, $r$ is restricted to $r \geq r_h$. The condition for holography to be internally consistent, that is, the (Lorentzian version of) the isoperimetric inequality (\ref{K}), is now far from trivial: it requires that $\mathfrak{S^{\m{L}}}(\m{AdSdyRN^{0}_{4}})(r)$ should be non-negative for all $r$. It turns out (see again \cite{kn:83}) that the condition for this is
\begin{equation}\label{M}
B^2\;+\;{\mu_B^2r_h^2\over L^4}\;\leq \;4\pi^3T^4,
\end{equation}
where the ``holographic dictionary" given earlier has been used, and where $r_h$ is to be understood as a certain function of $B$, $\mu_B$, and $T$, to be found by solving equation (\ref{J}). Thus \emph{the isoperimetric inequality in the bulk has a dual version on the boundary}: a complicated inequality to be satisfied by the boundary parameters $B$, $\mu_B$, and $T$.

If the inequality (\ref{M}) were not satisfied, then brane pair-production would run out of control, and the corresponding back-reaction on the bulk geometry would imply that it ceases to be static. Apart from the fact, mentioned earlier, that this would not be consistent, one should also consider that the dual interpretation would presumably be that the cosmic magnetic field, the temperature, and the baryonic chemical potential (or their ratios) would begin to evolve in a non-standard manner. But the evolution of these quantities with the scale factor is well-understood. Thus we should strive to ensure that (\ref{M}) holds in all cases.

One can show \cite{kn:83} that the inequality (\ref{M}), applied to equation (\ref{J}) above, is equivalent to the following inequality:
\begin{equation}\label{N}
\left (4\pi^3 - {8\alpha^2\over 9}\right )\varsigma_B^4\;+\;\alpha^3\varsigma_B \;-\;{3\alpha^4\over 4\pi}\;\leq \;0,
\end{equation}
where\footnote{Note that (\ref{M}) itself implies that $\beta^2 \leq 4\pi^3$, so there is no difficulty in the following definition of $\alpha$.}
\begin{equation}\label{O}
\alpha^2\;\equiv \;4\pi^3\;-\;\beta^2,
\end{equation}
and $\beta$ and $\varsigma_B$ are defined above in equations (\ref{E}) and (\ref{F}).

Fixing $\beta$ and regarding the left side of (\ref{N}) as a function of $\varsigma_B$, one finds that this function is increasing for all positive $\varsigma_B$, and that it has a single positive root. Therefore (\ref{N}) imposes an upper bound on $\varsigma_B$, given by this root, which we denote by $\rho(\beta)$; this is to be regarded as a known function, given by solving a quartic polynomial equation; one finds that it is in fact a decreasing function of $\beta$. Combining this statement with the inequality (\ref{B}) and recalling the definition of $\varsigma_B$, we have
\begin{equation}\label{P}
\mu_B^{CEP}/T^{CEP}\;\leq\;\mu_B^{\#}/T^{\#}\;\leq\;\rho(\beta).
\end{equation}
Thus, given a reasonable estimate for $\beta$, we can constrain $\mu_B^{\#}/T^{\#}$ (and the slope of the trajectory of the plasma through the phase diagram) from both sides.

\addtocounter{section}{1}
\section* {\large{\textsf{4. The Trajectory and the End of the Plasma Epoch}}}
The value of $\beta$ has been extensively investigated in theories of cosmic magnetogenesis \cite{kn:reviewA}\cite{kn:reviewB}. There is much uncertainty regarding its value during the plasma epoch, with estimates for $B$ at the end of the plasma epoch ranging up to around $3.7 \times 10^{17}$ gauss (though such a large value is very unlikely in the specific magnetogenesis mechanism given by Little Inflation). However, this means that $\beta$ only ranges from zero up to $\approx 1.15$, and, fortunately, it turns out that $\rho(\varsigma_B$) varies very little in this domain: a simple numerical investigation shows that $\rho(0) \approx 2.353$, while $\rho(1.15) \approx 2.324$. Let us take a value between these two extremes, say 2.34. Since we are taking $\mu_B^{CEP}/T^{CEP} = 300/145 \approx 2.07$, we can now express (\ref{P}) explicitly in the following manner:
\begin{equation}\label{Q}
2.07 \leq \mu_B^{\#}/T^{\#} \leq  2.34.
\end{equation}

Now we should strongly emphasise that, at this point at least, the gauge-gravity duality is not as precise as these numbers suggest. For, as is well known, it applies to boundary field theories which resemble strongly coupled QCD in many ways, and which share surprisingly many universal features with it, yet are not the same. The generally accepted procedure (see \cite{kn:karch} for a detailed discussion) is that gauge-gravity predictions should be interpreted as accurate to within a factor of about 2 (systematic error). A reasonably realistic interpretation of (\ref{Q}) might state that $\mu_B^{\#}/T^{\#}$ is constrained to lie between approximately 2 and 4.

Note that Little Inflation itself allows for far larger values, up to 100 (see \cite{kn:tillmann3}); thus, despite its limitations, the gauge-gravity duality makes a useful contribution here, narrowing the range of possible values for $\mu_B^{\#}/T^{\#}$ very substantially. In fact, the range is narrower than $2 \leq \mu_B^{\#}/T^{\#} \leq 4$: one can argue that both the lower and the upper ends of the range must be avoided. 

To begin with the lower limit: as with any critical point, the immediate vicinity of the quark matter critical endpoint is beset by strong fluctuations: this is the well-known phenomenon of \emph{critical opalescence} \cite{kn:csorgo}, the detection of which is in fact the precise goal of experiments relevant to this regime \cite{kn:STARCRIT,kn:PHENIX,kn:turko}. It is highly likely that strong fluctuations in this vicinity would disturb the establishment and persistence of a metastable QCD vacuum, which is the very basis of Little Inflation. Therefore we argue that, in order for Little Inflation to be viable, values of $\mu_B^{\#}/T^{\#}$ around 2 must be avoided.

On the other hand, as we have already mentioned, the upper bound set by gauge-gravity duality is uncertain, by the very nature of the duality in its present state of development. The existence of a stable plasma violating the isoperimetric inequality (\ref{K}) would prove that gauge-gravity duality cannot be consistently applied to the cosmic QGP. Thus it would be risky to entertain values for $\mu_B^{\#}/T^{\#}$ near to 4.

To be concrete, let us take $\mu_B^{\#}/T^{\#} = 3$ as a reasonable estimate consistent with all of the above considerations. The trajectory is then a line in the phase diagram of slope 1/3; one finds, using the parametrization of the phase line described in \cite{kn:karsch}, that the intersection with the phase line is at approximately ($\mu_B^{\#},\,T^{\#}$) = (400, 135) in units of MeV. This is at an acceptable distance from our assumed location of the critical endpoint, ($\mu_B^{CEP},\,T^{CEP}$) = (300, 145), so that fluctuations should not disrupt Little Inflation, while remaining compatible with the requirements of gauge-gravity duality.

\addtocounter{section}{1}
\section* {\large{\textsf{5. Conclusion: Initial Conditions for the Hadronic Epoch}}}
To summarize: using the latest lattice results and methods drawn from gauge-gravity duality, we have argued that, in the highly interesting and novel Little Inflation theory, the following statements can be made.

$\bullet$ The trajectory of the cosmic plasma through the quark matter phase diagram is a straight line (at least in the late plasma epoch) such that $2 \leq \mu_B^{\#}/T^{\#} \leq 4$, though the ends of this range should be avoided;

$\bullet$ If we take $\mu_B^{\#}/T^{\#} \approx 3$, and use the parametrization given in \cite{kn:karsch}, the end of the plasma epoch occurred near to the point in the diagram with coordinates ($\mu_B^{\#},\,T^{\#}$) = (400, 135) MeV (assuming that the critical end point is near (300, 145) MeV); that is, at a temperature not far below that of the critical end point, but at baryonic chemical potential substantially higher.

It follows that the beam scan experiments \cite{kn:ilya,kn:dong,kn:STAR,kn:shine,kn:nica,kn:fair,kn:BEAM}, many of which may ultimately be capable of probing this regime, will, under these assumptions, be examining a plasma similar to the cosmic plasma just before it underwent a first-order phase transition. The prospect of having direct access to matter akin to the content of the microsecond Universe, just as it was undergoing a strong phase transition, is exciting indeed.

\addtocounter{section}{1}
\section*{\large{\textsf{Acknowledgements}}}
The author is grateful to Prof Soon Wanmei for technical assistance, and to Cate Yawen McInnes for encouragement.

\end{document}